\documentclass[aps,preprintnumbers,prl,twocolumn,superscriptaddress]{revtex4}
\usepackage{epsfig,latexsym,cancel,amssymb,amsmath}
\usepackage{graphicx}
\usepackage{color}
\usepackage[usenames,dvipsnames,svgnames]{xcolor}
\usepackage{hyperref}
\usepackage[normalem]{ulem}
\usepackage{framed}
\usepackage{dcolumn}
\usepackage{float}
\usepackage{verbatim}
\usepackage{pdflscape}
\usepackage{array}
\usepackage[caption=false]{subfig}  
\newcolumntype{x}[1]{>{\centering\let\newline\\\arraybackslash\hspace{0pt}}p{#1}}

\def\beq{\begin{equation}}
\def\eeq{\end{equation}}
\def\bea{\begin{eqnarray}}
\def\eea{\end{eqnarray}}

\begin{document}

\title{Observationally inferred dark matter phase-space distribution and direct detection experiments}

\author{Sayan Mandal}
\affiliation{McWilliams Center for Cosmology and Department of Physics, Carnegie Mellon University, 5000 Forbes Ave, Pittsburgh, PA 15213, USA}
\affiliation{Department of Theoretical Physics, Tata Institute of Fundamental Research, Homi Bhabha Road, Mumbai 400005, India}
\author{Subhabrata Majumdar}
\affiliation{Department of Theoretical Physics, Tata Institute of Fundamental Research, Homi Bhabha Road, Mumbai 400005, India}
\author{Vikram Rentala}
\affiliation{Department of Physics, Indian Institute of Technology Bombay, Powai, Mumbai 400076, India}
\author{Ritoban Basu Thakur}
\affiliation{Kavli Institute for Cosmological Physics, University of Chicago, 5640 S. Ellis Ave., Chicago, IL 60637, USA}

\begin{abstract}

We present a detailed analysis of the effect of an observationally determined dark matter (DM) velocity distribution function (VDF) of the Milky Way (MW) on DM direct detection rates. We go beyond local kinematic tracers and use rotation curve data up to 200 kpc to construct a MW mass model and self-consistently determine the local phase-space distribution of DM. This approach mitigates any incomplete understanding of local dark matter-visible matter degeneracies that can affect the determination of the VDF. Comparing with the oft used Standard Halo Model (SHM), which assumes an isothermal VDF, we look at how the tail of the empirically determined VDF alters our interpretation of the present direct detection WIMP DM cross section exclusion limits.
While previous studies have suggested a very large difference (of more than an order of magnitude) in the bounds at low DM masses, we show that accounting for the detector response at low threshold energies, the difference is still significant although less extreme.
The change in the number of signal events, when using the empirically determined DM VDF in contrast to the SHM VDF, is most prominent for low DM masses for which the shape of the recoil energy spectrum depends sensitively on the detector threshold energy as well as detector response near the threshold. We demonstrate that these trends carry over to the respective DM exclusion limits, modulo detailed understanding of the experimental backgrounds.
With the unprecedented precision of astrometric data in the GAIA era, use of observationally determined DM phase-space will become a critical and necessary ingredient for DM searches.
We provide an accurate fit to the current best observationally determined DM VDF (and self-consistent local DM density) for use in analyzing current DM direct detection data by the experimental community.

\end{abstract}

\maketitle
\section{Introduction}
One of the most popular candidates for the dark matter (DM) particle is the hypothetical \textit{weakly interacting massive particle} (WIMP) \citep{Kamionkowski:1997zb, JUNGMAN1996195, Patrick-Lux,  Ferrer2006, BertoneDetection2005, Davis2015}. The determination of the density of these particles in the halo of the Milky Way (MW)  galaxy in general and the solar neighborhood in particular is crucial for many direct detection experiments which attempt to measure the rate of nuclear recoil events caused by the WIMPs scattering off of the detector target nuclei. The expected scattering behavior is strongly dependent on the local astrophysical properties of DM. In particular, the scattering rate is directly proportional to the local DM density. In addition, the velocity distribution function (VDF) of the DM crucially affects the shape of the nuclear recoil energy spectrum. Thus, it is imperative to have precise knowledge of the local phase-space distribution of DM from observations in order to set precise bounds on the DM particle physics parameter space~\citep{Annika2011}.

DM direct detection experiments usually assume the simplest possible `Standard Halo Model' (SHM)
for the DM halo, in which the velocity distribution is Maxwellian. This model assumes the
halo to be an isotropic, isothermal sphere - hypotheses that are unlikely to be valid in reality.
Moreover, N-body simulations produce halos with
velocity distributions which deviate systematically from a
Maxwellian \citep{Kuhlen:2009vh,Ling:2009eh,Sparre:2012zk,Kuhlen:2013tra, Butsky:2015pya,Bozorgnia:2016ogo}.
One can also construct more realistic, analytical VDFs that differ from the predictions of the SHM \citep{Widrow:2000, BinneyTremaine:2008,Evans:2006,Eilersen:2017}. However, a self consistent connection to observations is generally missing - for example, simulations obtain a VDF of a Milky Way `like' halo identified inside a simulation box using user defined criteria (such as halo mass or circular velocity); on the other hand, analytic models typically cannot incorporate the effect of the baryonic mass component of the MW on the DM VDF.

In this work, we advocate for a more realistic, observationally-driven  approach which follows a three step procedure - (i) first, visible matter tracers in the Milky Way are used to map out the  gravitational potential, $\Phi(r)$, in the region of interest, (ii) second, a multi-component mass density model, having both DM and visible matter (VM) in various configurations is obtained consistent with the observed gravitational potential, and (iii) finally, since the density is an integral of the total phase-space distribution function, one can invert the equation connecting the two to obtain the VDF. A convenient way to get the inversion is to use the Eddington formalism \cite{BinneyTremaine:2008}. The full procedure results in a self-consistent determination of both the DM density and its velocity distribution. This was first done by \citet{Bhattacharjee:2012xm}; this work is based on similar, but more detailed, MW DM phase-space analysis using a larger dataset \citep{BhattRotCurv}.

Our approach, of using rotation curve (RC) data up to $\sim 200$ kpc as the visible matter tracer, makes us sensitive to the DM and VM distributions throughout the halo. In this approach it is easier to separate the contributions of the different components, modulo the analytic form of the DM distribution that is assumed \footnote{We use a \textit{Navarro-Frenk-White} (NFW) \cite{1996ApJ...462..563N} DM profile. However, we have found that the determination of the local DM phase-space is not very sensitive to other choices of the galactic DM profile.}.  Estimates of the local DM phase-space based on local dynamics rely on the accuracy of separating the VM contributions and are prone to contaminations, viz.  without the leverage of a large number of galactic radial bins, a Bayesian analysis marginalizing over the VM contribution using only local data has large degeneracies. Our method of using data at a large number of radial points leads to substantially better breaking of the degeneracy between the VM and DM distributions.

For simplicity, experimental DM direct detection exclusion curves in the literature are almost always calculated by assuming a na\"{i}ve SHM expectation for the MW DM halo.
There have been a number of theory papers that have tried to go beyond the SHM and assess the impact of astrophysical uncertainties on DM direct detection experiments (for example, see \cite{Frandsen2012, Fairbairn2013}). Some of these attempts have examined the impact of an uncertain local dark matter density (such as \cite{Green2012}), which results in a trivial re-scaling of the exclusion curves. Others have attempted to look at the effect of the DM velocities via changes in the local DM escape speed and the local DM velocity dispersion (see \cite{McCabe2010}). Studies that tried to incorporate the full local DM VDF, other than SHM, have been mainly restricted to ansatzes or VDFs extracted from simulations \cite{McCabe2010, Frandsen2012, Mao2013}.

We claim that the right approach should be to use an {\it observationally inferred} determination, along with the related uncertainties, of the local DM phase-space in a {\it self-consistent} manner (also see \footnote{An earlier work by \citet{CatenaUllio2012} used Eddington's
method to derive the local VDF from various dynamical constraints on the gross properties of the Galaxy rather than the full RC data as done here. There has also been a recent study using the local kinematics of stars to estimate the local DM VDF \cite{Herzog2017b}; however, similar to \cite{Bhattacharjee:2012xm}, the implications of the inferred VDF on DM direct detection, as compared to using the SHM, was simplistically taken to be given by only the ratio of the corresponding velocity integrals (described in detail later in this work).}). In this work, we take a detailed look at this fundamentally important element of  DM detection results. The main result of our work is the first re-estimation the DM exclusion curves, for some of the major DM direct detection  experiments, using  observationally determined local DM phase-space.

\section{Self-consistent determination of the DM phase-space distribution}
The rate of nuclear recoil events, in direct detection searches, depends  crucially on the local (i.e. solar neighborhood) density and velocity distribution of the WIMPs in the Galaxy, which are a priori unknown. In contrast to the density, not much knowledge directly based on observational data is available on the likely form of the velocity distribution function (VDF) of the WIMPs in the Galaxy. The standard practice is to use what is often referred to as the Standard Halo Model (SHM), in which the DM halo of the Galaxy is described as a single-component isothermal sphere, for which the VDF is assumed to be isotropic and of the Maxwell-Boltzmann form. High resolution cosmological simulations of DM halos  give strong indications of significant departure of the VDF from the Maxwellian. On the other hand, these cosmological simulations do not yet satisfactorily include the gravitational effects of the visible matter components of the real galaxy, namely, the central bulge and the disk.

One approach to determining the local density of DM is to use the rotation curve data to find the likelihood of the parameters characterizing the density distributions of the various mass components of the galaxy. In general, the visible matter (VM) parameters are fixed (from observational data) and the dark matter (DM) parameters are obtained from a likelihood maximization. A full likelihood analysis (DM and VM) was first done \citep{Bhattacharjee:2012xm} by taking the \textit{Navarro-Frenk-White} (NFW) profile \citep{1996ApJ...462..563N} for the DM halo, a spheroidal bulge and an axisymmetric disk \citep{CaldwellModels,Kuijken1,Kuijken2,Kuijken3,Kuijken4}. The NFW DM density profile is given by $\rho_\mathrm{DM}(r)=\rho_{\mathrm{DM},\odot}\left(\frac{R_\odot}{r}\right)\left(\frac{r_s+R_\odot}{r_s+r}\right)^2$ where $r$ is the distance from the galactic center, $\rho_{\mathrm{DM},\odot}$ is the local DM density, $r_s$ the scale radius of the halo and $R_\odot$ is the distance of the sun from the galactic center. The bulge and disk density profiles are given respectively by $\rho_b(r)=\rho_{b0}\left(1+\frac{r^2}{r_b^2}\right)^{-\frac{3}{2}}$ and  $\rho_d(R,z)=\frac{\Sigma_\odot}{2z_d}\exp\left(-\frac{R-R_\odot}{R_d}-\frac{|z|}{z_d}\right)$ (in cylindrical coordinates), where $\rho_{b0}$  is the normalization of the bulge density, $\Sigma_\odot$ is the local disk surface density, and $r_b$ and $R_d$ are the scale radii of the bulge and the disk respectively. The parameter $z_d$ is the scale height of the disk. The visible matter parameterizations are based on fits to local kinematical data. This fiducial model of the MW consisting of a dark matter halo, visible matter bulge and a single disk is a minimal model of the mass distribution in the MW  \footnote{We have checked that the simple extension of adding extra visible matter mass components (like a thin plus a thick disk or a central blackhole) does not not change the dark halo parameters appreciably, while largely altering the best fit visible matter parameters.}.

For a given choice of the density profiles of both DM and VM, we can use the Poisson equation  to obtain the total gravitational potential $\Phi(R)$ at a given radius on the galactic plane, and thus we get the circular rotation speed at $R$ from the relation,
\begin{equation}\label{Eq1}
v_c^2(R)\,=\, - R\frac{\partial}{\partial R}\Phi(R,z=0).
\end{equation}
Next, a \textit{Markov Chain Monte Carlo} (MCMC) analysis is carried out to determine the most likely values of the density parameters (along with their $1\sigma$ uncertainties). In this work, we adopt the \textit{Python} framework \texttt{CosmoHammer} \citep{Akeret201327} which embeds the \texttt{emcee} package by Foreman-Mackey et. al. \cite{foremanmackey2013} that is based on an improved MCMC algorithm by Goodman and Weare \citep{goodweare}.  The $\chi^2$ test statistic used within the MCMC is $\chi^2=\sum_{i=1}^N\left(\frac{v_{\text{obs},i}-v_{\text{theo},i}}{\sigma_i}\right)^2$, where $v_{\text{obs},i}$ is the observed circular velocity value, $v_{\text{theo},i}$ is the one theoretically calculated, $\sigma_i$ the error in the observed velocity value  and $N$ is the total number of binned data points at different distances from the galactic center. The
 best fit density parameters are used to estimate the full spatial density of the DM particles $\rho_\mathrm{DM}(\mathbf{r})$ and the \textit{total} gravitational potential $\Phi(\mathbf{r})$.

Under isotropic conditions , the phase space distribution function $\mathcal{F}$ of the DM component, at a position $\mathbf{r}$, depends only on the total \textit{specific} energy $E=\frac{1}{2}v^2+\Phi(r)$, with $v=|\mathbf{v}|$, $r=|\mathbf{r}|$. This function $\mathcal{F}$ can be uniquely determined using the \textit{Eddington formula} \citep{BinneyBook},
\begin{equation}\label{eCurlyF}
\mathcal{F}(\mathcal{E})=\frac{1}{\sqrt{8}\pi^2}\left[\int_0^\mathcal{E}\frac{d\Psi}{\sqrt{\mathcal{E}-\Psi}}\frac{d^2\rho}{d\Psi^2}+\frac{1}{\sqrt{\mathcal{E}}}\left(\frac{d\rho}{d\Psi}\right)_{\Psi=0}\right]
\end{equation}
where $\Psi(r)\equiv-\Phi(r)+\Phi(r=\infty)$, $\mathcal{E}\equiv-E+\Phi(r=\infty)=\Psi(r)-\frac{1}{2}v^2$, and $\rho(r)$ is the total density. The VDF at a radius $r$ can be obtained as,
\begin{equation}\label{efrv}
f_{\mathbf r}(\mathbf{v})=\frac{\mathcal{F}}{\rho(r)}.
\end{equation}
Also, for $\mathcal{E}>0$, $\mathcal{F}>0$, and for $\mathcal{E}<0$, $\mathcal{F}=0$; this ensures that the VDF truncates naturally at the escape velocity $v_\mathrm{esc}=\sqrt{2|\Psi(r)|}$.
We work with the \textit{normalized, one-dimensional} velocity distribution function $f_r(v)\equiv 4\pi v^2 f_{\mathbf r}(\mathbf{v})$, which satisfies $\int_0^{v_\mathrm{esc}}f_r(v)\,dv=1$.

A key assumption in using Eddington's method is that of isotropy (i.e, the net potential has spherical symmetry).
Due to the axisymmetric
nature of the VM disk, the total potential in our MW mass model is non-spherical. In order to use the Eddington approximation, we use a spherical approximation \cite{CatenaUllio2012} for the VM potential given by
$\Phi_{VM}(r)\,\simeq \, \int_0^r M_{\rm VM}(r^\prime)/{r^\prime}^2 dr^\prime$, where $M_{\rm VM}(r)$ is the mass of the VM within a radius $r$.

The DM density $\rho_{DM}(r)$ is obtained by
integrating the single-particle phase-space distribution function (which in the simplest case
is a function of total energy per unit mass of a DM particle) over the
velocities of the DM particle. Under the spherical approximation, $\rho_{DM}(r) = \int \mathcal{F}[E=(1/2)v^2+\Phi(r)] 4\pi v^2 dv$ where $v$ is the DM velocity and $\mathcal{F}$ is the phase-space distribution function, which is inverted give the DM VDF. Note that $\rho_{DM}$ is also implicitly present on the right hand side of the above equation in $\Phi(r)$, and hence the equation demands
a self-consistent solution. For a given DM and VM distribution that is consistent with the particular $\Phi(r)$ `at all $r$', a unique solution for the VDF  $f(r,v)\equiv
\mathcal{F(E)}/\rho_{DM}(r)$ `at all $r$' exists in the spherically symmetric case.

The VDF (in eq. \eqref{efrv}) determined using the technique outlined in the preceding discussion is completely empirical, without any reference to any simulations; it is connected to the local DM density in a self-consistent manner through the Eddington formula and we only need to model the potential contribution from the disk as arising due to a spherical mass distribution. We find that the spherical approximation induces corrections of the order of $\approx 10\%$ in the value of $v^2(R_\odot)$).

We stress that the arguments presented in this section apply to virialized DM halos. Kinematic outliers associated with local DM substructure such as streams or debris flows \citep{Diemand:2008in, abcdef, Freese04Str, Savage:2007zz, Kuhlen970, KuhlenEtal2012, Lisanti2012, Lisanti2015} could also impact the interpretation of the experimental results. Because the origin of these outliers is unknown, however, we choose to not include them. Additionally, recent mergers of satellite galaxies could lead to spatial or kinematic substructure. We also ignore the possibility of velocity spikes in the VDF due to local substructure~\citep{Kuhlen:2009vh}.

\section{Velocity profiles}

Many authors have determined the RC of the Galaxy, using kinematical and positional information
for some tracer objects moving in the gravitational potential of the galaxy.
In general, one measures line-of-sight (LOS) quantities (positional and kinematical data)
and the RC is derived from these.
To determine the RC for the disk region, one has to adopt a value for the the
\textit{local standard of rest} (LSR) which corresponds to the position $(R_0)$ and
velocity $(v_{c0})$ of the sun with respect to the galactic center, and make the assumption
that the tracer objects follow a circular orbit around the galactic center.
From this, the positions and velocities about the galactic center can be obtained.
The choice of LSR plays a crucial role in determining the parameters of the mass
model of the MW - it affects the value of local DM density and our estimation of
other MW properties, like the mass of the MW.
For our analysis, we follow \citep{Bhattacharjee:2012xm, MandalMajumdar:2018a},
and pick two popular LSRs used in MW studies: (i) $R_0 = 8$ kpc and  $v_{c0} = 200$ km/s
and (ii) $R_0 = 8.5$ kpc and  $v_{c0} = 220$ km/s.

The various tracer objects used for constructing the RC in \citep{BhattRotCurv} include HI and HII regions
(CO emissions from the latter), Cepheids, Planetary nebulae, etc.
For regions extending beyond the visible disk of the galaxy, they look at tracers
distributed in the halo of the Milky Way (like dwarf spheroidals, globular clusters, K-giant stars, etc.).
The latter tracers are of the non-disk kind, and their motion around the galactic center is typically unsystematic.
Under the assumption that these objects are isotropically distributed in the halo, one can
define an effective circular velocity $v_c$ at a galactocentric distance $r$, and use the Jeans equation
to relate it to the observed number densities and velocity dispersions \cite{BinneyBook, BhattRotCurv}.
One disadvantage of using these non-disk tracers is that they can have non-negligible velocity anisotropy.
Currently, only the line-of-sight velocity dispersion is known to precision  at large distances, the
velocity anisotropy cannot be well determined \footnote{Precision astrometric measurements, such
as with GAIA, will lead to better determination of the velocity anisotropy.}.
In this work, we neglect the effect of any such anisotropy in the RC data.
For the effect of velocity anisotropy on the RC at large distances, we refer the
reader to \citep{CatenaUllio2010, BhattRotCurv, Rashkov2013}.

After the raw RC is generated as above, it is suitably radially binned and averaged.
The binning strategy used by Bhattacharjee et.al. \citep{BhattRotCurv} is two-fold.
In the first step, they average over each individual data set, choosing different bin
sizes at different radius ranges (smaller bins at low $r$ where there is more data, etc).
These bin sizes are manually optimized to best reflect the overall behavior of the raw data points.
Once these individual datasets are binned and averaged, they are compiled into a larger
dataset, and the above process is repeated once again to arrive at the final dataset.

We performed an MCMC analysis on the RC compiled by \citet{BhattRotCurv}
following the detailed procedure described above, to obtain the best fitting
DM plus VM density distributions of our galaxy.
There is a wealth of knowledge available on the distribution of visible matter in
the MW based on decades of astrophysical observations, and it is prudent to
add some of this information in the form of VM priors in our MCMC analysis.
In this Bayesian approach, the final best fit DM distribution depends on the
imposed priors on the VM distribution.
Although, we have focussed on RC, which gives us a estimate of the gravitation potential over a large radial range, one can consider other constraints
such as the vertical force in the solar neighbourhood and the Oort constants in any disk-bulge-halo models \cite{DehnenBinney98, WidrowDubinsky05, Widrowetal08}. A detailed discussion of
local kinematics and comparison with RC inferred parameters is presented elsewhere \citep{MandalMajumdar:2018a}.

Using the Eddington formula, we estimated the local DM VDF, which is self-consistently related to the DM and VM mass distribution, and in particular to the DM local density. Note, that the extracted DM VDF implicitly depends on the choice of VM prior and LSR.
Although flat priors give the most `unbiased' best fit parameters, we use the wealth of knowledge on the VM distribution to impose comparatively tighter  local VM priors. For all the results quoted in the rest of the paper, we chose our priors based on observational constraints on the local VM density \citep{Kuijken2,refId0,Sivertsson:2017rkp,Zhang13}, which can be expressed in terms of constraints on the disk parameters $\Sigma_\odot$ and $R_d$. For our analysis, we adopt the following \textit{gaussian} priors on the disk parameters: $\Sigma_\odot=67\pm8\,\mathrm{M}_\odot\,\mathrm{pc}^{-2}$ and $R_d=2.3\pm0.6\,\mathrm{kpc}$. Singh et al. \citep{MandalMajumdar:2018a} have studied the impact of the choice of local VM prior and LSR on the estimates of the MW DM distribution and, in this work, we adopt their method and results.

\begin{figure}[htp]
\centering
\subfloat[{Rotation curve data from Bhattacharjee et al.~\cite{BhattRotCurv} along with the MCMC best fit (blue). Also shown is the decomposition of the RC into visible Disk + Bulge (green) and DM halo (orange) components.}]
{%
  \includegraphics[clip,width=1\columnwidth]{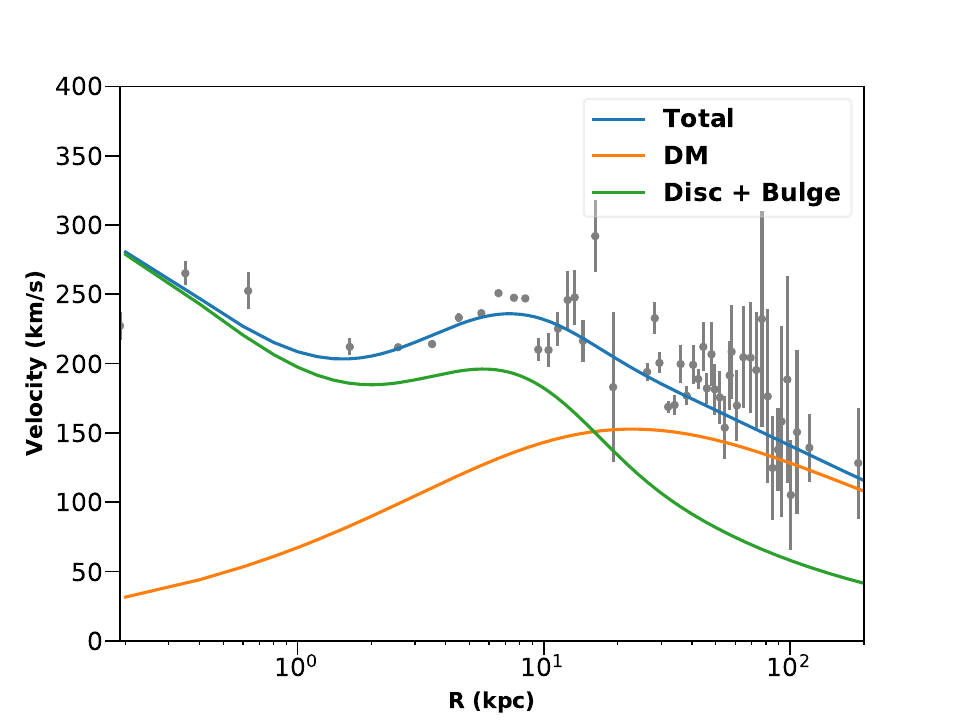}%
\label{fig:rc}
}

\subfloat[MCMC correlation between the DM scale radius $r_s$ and the local DM density $\rho_{DM,\odot}$, the shaded contours show the 1, 2 and 3-$\sigma$ confidence regions.]{%
\hspace*{-1cm}
  \includegraphics[clip,width=0.85\columnwidth]{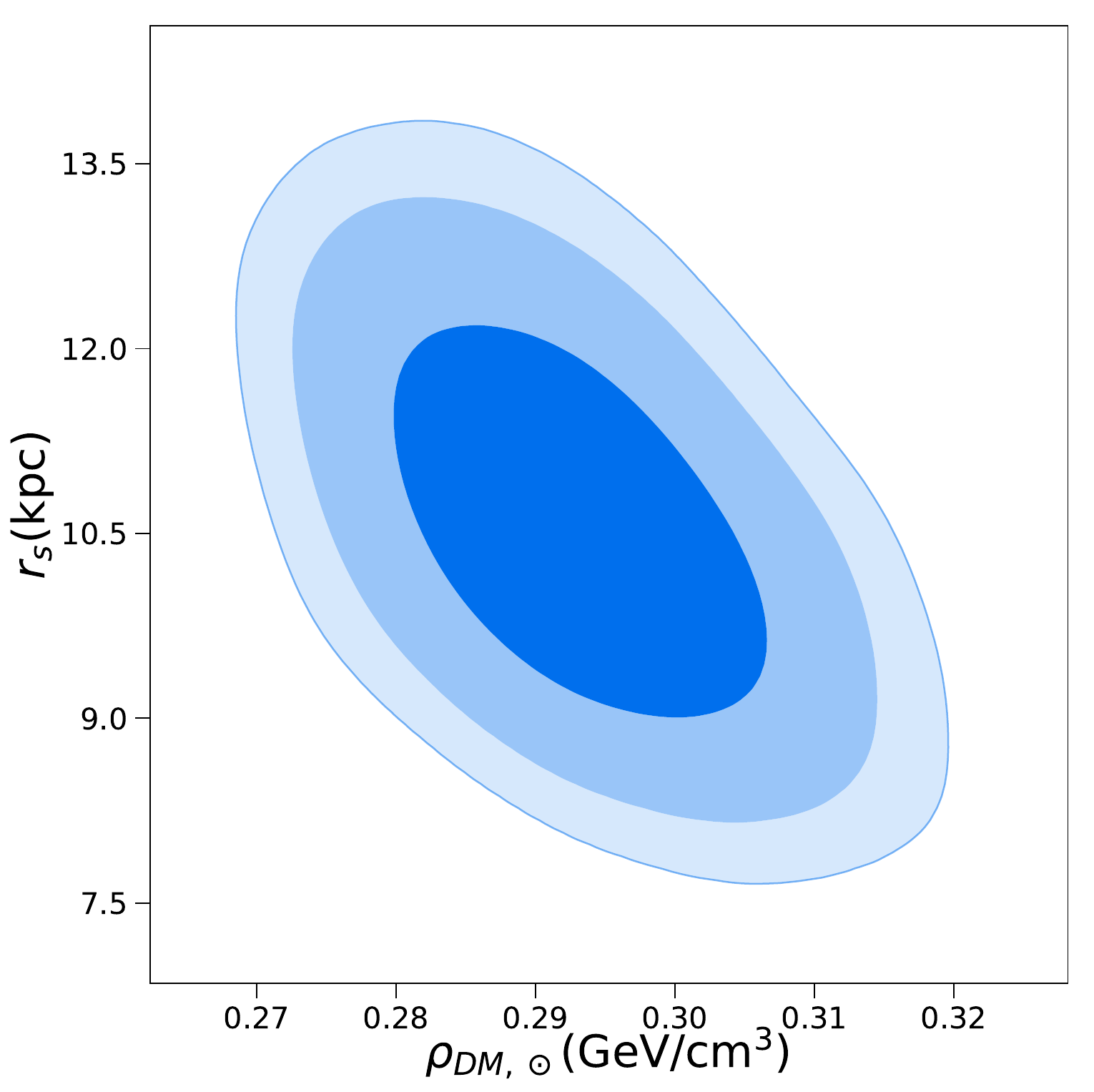}%
\label{fig:rhodmcorr}
}
\\
\subfloat[MCMC correlation between the local VM disk surface density $\Sigma_0$ and the local DM density $\rho_{DM,\odot}$, the shaded contours show the 1, 2 and 3-$\sigma$ confidence regions.]{%
\hspace*{-1cm}
  \includegraphics[clip,width=0.85\columnwidth]{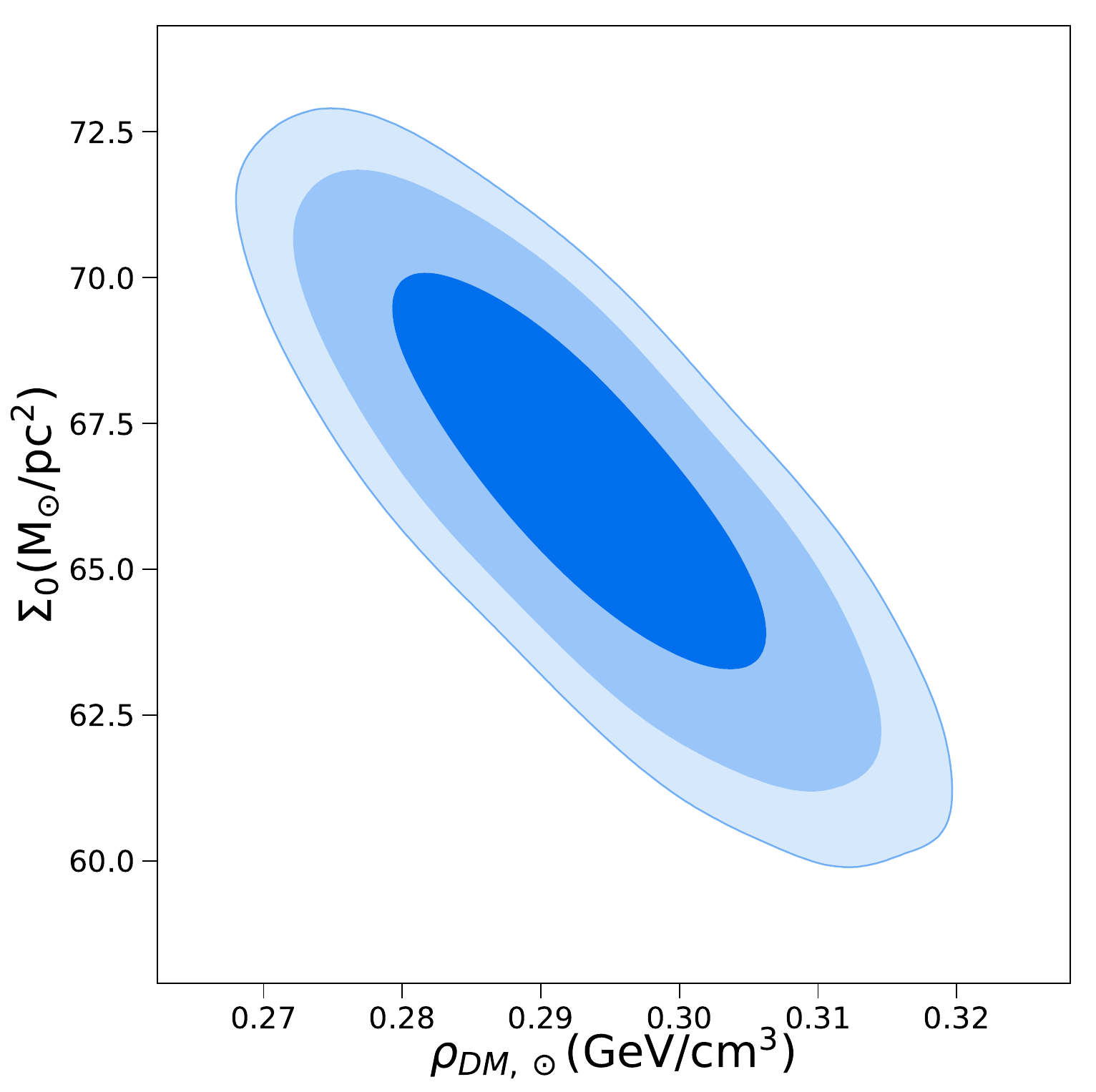}%
 \label{fig:rhodmcorr2}
}

\caption{Results for MCMC fit to the rotation curve data.}

\end{figure}

\begin{figure}
\includegraphics[width=8cm]{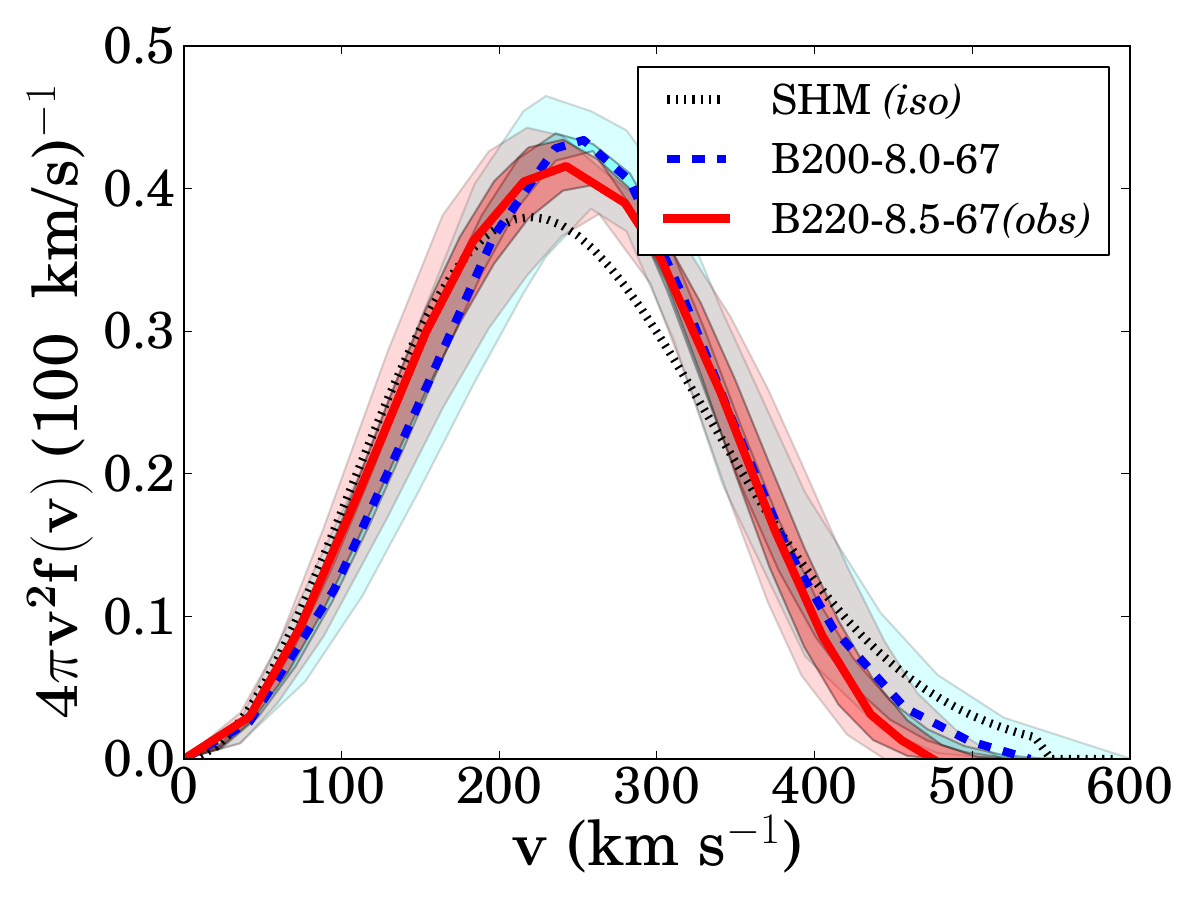}
\caption{Plot of the observationally determined velocity profiles in our local neighborhood. The solid red and dashed blue curves correspond to the observationally determined VDFs with different choices of the local standard of rest. The thick and thin error bands around each show the error envelope assuming \textit{current} errors on the rotation curve data as well as projected \textit{third} errors, respectively. The canonical isothermal/Maxwellian VDF is shown by the thin black dotted line (SHM). The two observationally inferred VDFs are non-Maxwellian in nature and differ from the SHM VDF at both low and high velocity tails. The departure of the observed VDFs at high velocity tails crucially impacts the interpretation of low mass dark matter searches in direct detection experiments.}
\label{fig:profile}
\end{figure}

Together with our choice of VM prior and two sets of LSRs, we determine the corresponding local DM densities and VDFs:
\begin{enumerate}

\item {\bf B220-8.5-67}: For the choice of LSR with $R_0 = 8.5$ kpc and $v_{c0} = 220$ km/s, we show the rotation curve data from \citep{BhattRotCurv} along with the best fit total, DM, bulge and disc RC decomposition from our MCMC analysis in Fig.~\ref{fig:rc}. We find $\rho_{DM,\odot}=0.29\pm0.02 \,\mathrm{GeV \,}{\rm cm}^{-3}$. In Figs.~\ref{fig:rhodmcorr} and ~\ref{fig:rhodmcorr2}, we show two of the main MCMC correlations for the local DM density $\rho_{DM,\odot}$; viz. with the DM scale radius $r_s$ and the local visible matter disk density $\Sigma_\odot$, respectively. The full set of MCMC correlations will be presented in a forthcoming work \citep{MandalMajumdar:2018a} which will more exhaustively examine multiple choices of VM priors, LSRs, datasets (pre and post-GAIA RC and local kinematics) etc.
\\
The corresponding VDF is shown in Fig.~\ref{fig:profile} with the high velocity cutoff corresponding to the local escape velocity which we find to be $v_{\textrm{esc}}=475.00\,\mathrm{km\,s^{-1}}$.

\item {\bf B200-8.0-67}: For the choice of LSR with $R_0 = 8$ kpc and  $v_{c0} = 200$ km/s, we find the best fit local DM density to be $\rho_{DM,\odot}=0.18\pm0.02 \,\mathrm{GeV}{\rm cm}^{-3}$. Again, the VDF is shown in Fig.~\ref{fig:profile}. The local escape velocity with this LSR is found to be $v_{\textrm{esc}}=536.83\,\mathrm{km\,s^{-1}}$.
\end{enumerate}


In the rest of the paper, the observationally determined VDF B220-8.5-67 is selected as our fiducial VDF and we denote it as \textit{obs} for the sake of brevity. From Fig.~\ref{fig:profile}, we can see that the  VDF B200-8.0-67  is very close to our fiducial VDF and the main difference between the expected DM signal rates for the two choices of LSR enters through a trivial rescaling of the rate due to the different best-fit local DM densities.

The fractional uncertainty on the local DM density (as listed above) is between $5-10\%$ irrespective of the choice of LSR or any observational priors on the VM. In contrast, estimates of $\rho_{DM,\odot}$ using local kinematics typically has an uncertainty of $\sim 30\%$ \cite{BovyTremaine2012}. We attribute this difference to our use of data at a larger number of radial bins which leads to the breaking of any DM-VM degeneracies. In the rest of the paper, we neglect this small uncertainty on $\rho_{DM,\odot}$ since the VDF uncertainty becomes the dominant astrophysical uncertainty for DM direct detection searches.

Note, that the VDF is determined from the mean values of the determined model parameters, and the 1-$\sigma$ errors on these parameters gives us an uncertainty band on the VDF. The best fit parameters of the MW mass model have correlated uncertainties and these correlations are naturally computed in the MCMC analysis. These correlations lead to correlations in errors on the VDF at different velocities \footnote{The uncertainty band on the VDF is the envelope of all the VDFs obtained by varying the mass model parameters, one at a time, at the $1\sigma$ level. Due to the correlations in errors on the extracted parameters, this envelope is generally an overestimate of the $1\sigma$ uncertainty on the VDF at each velocity.}. In our analysis, we use the above RC datasets with the error bars on the velocities (a) having the reported values (which we call \textit{current}), and (b) reduced by 1/3 (which we call \textit{third}). The reduction of the error bars on the RC unsurprisingly leads to the narrowing of the error bands on the VDF. Our analysis with the reduced error bars models upcoming data from the GAIA satellite \cite{Gaia2016}. Compared to the $\sim 50$ RC data points \cite{BhattRotCurv} that we have used in this work,  GAIA data will potentially have an order of magnitude larger number of RC data points within 200 kpc which will reduce the relative error on the velocity by a factor of $\frac{\sigma_v}{v}\sim\frac{1}{\sqrt{N}}\sim \frac{1}{3}$, where $N=10$ is a benchmark increase in the number of tracers that we expect, thus giving tighter constraints on the VDF and the local DM density.

\begin{figure}
\includegraphics[width=8.5cm]{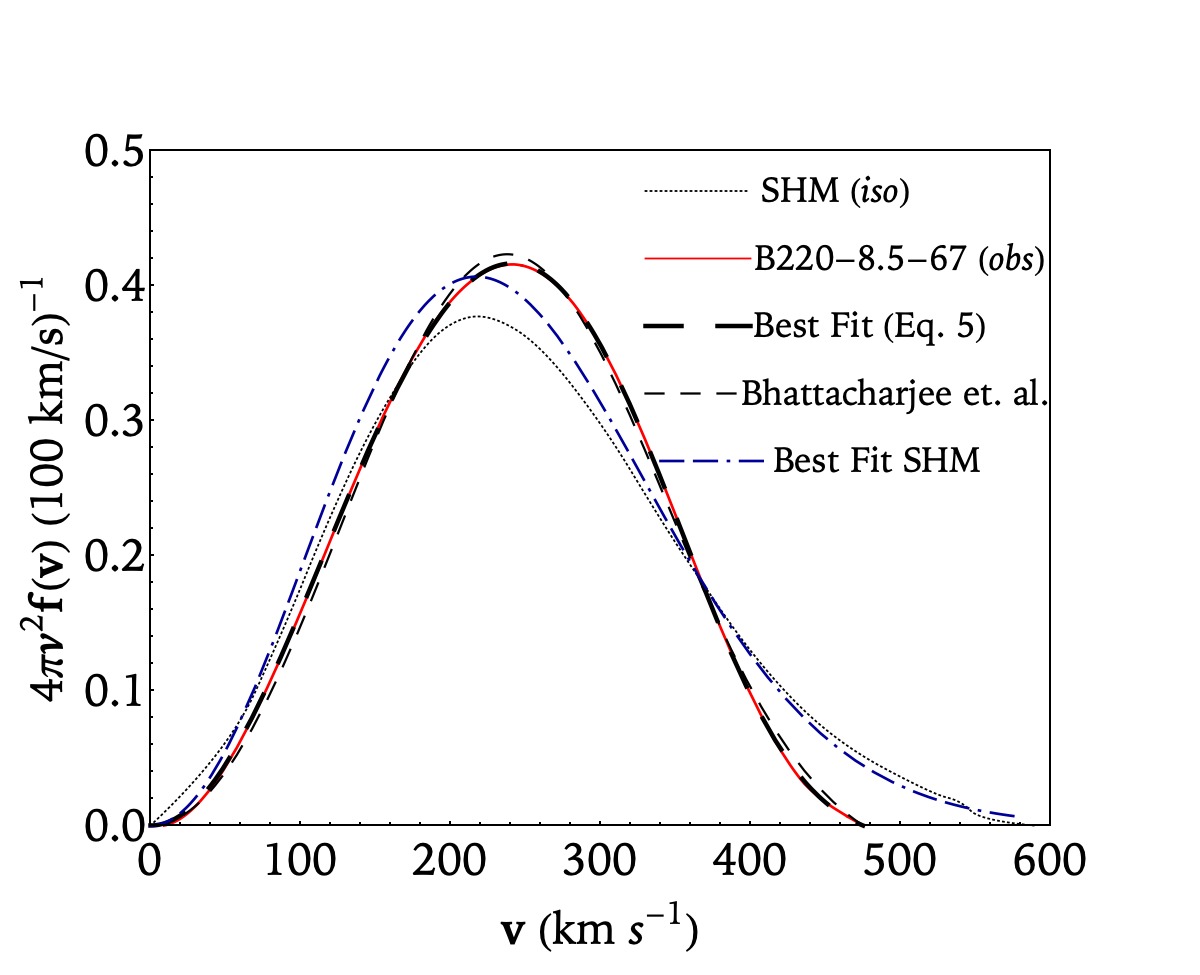}
\caption{The {\it obs} VDF (solid red line) is shown along with the best fit (thick black dashed line) using the fitting function in eq.~\eqref{Fit2}. The fit agrees to within 1\% over almost the entire velocity range. For comparison the best fit using the fitting function given in \cite{Bhattacharjee:2012xm} (thin black dashed line), and the SHM {\it iso} VDF (dotted black line) are also shown. We have also shown the best fit Maxwellian to the {\it obs} VDF (dash-dot blue line).
 }
\label{fig:Fit85}
\end{figure}

As a benchmark for comparison, we introduce the {\bf SHM} which is conventionally used in all direct detection DM experimental analyses.
The VDF is {\it assumed} to be given by a Maxwellian distribution. Note, that the corresponding self-consistent DM density distribution assumes that the DM halo is a singular isothermal sphere. Moreover, the SHM assumes a single component halo mass model, an assumption which breaks down in the presence of VM.  The isothermal velocity profile of dark matter as a function of the dark matter velocity in the galactic rest frame $v$ is given by,
\begin{equation}\label{Maxw}
f_\textrm{iso}(\mathbf{v}) = k_\textrm{iso} \exp{(-v^2/v_d^2)},
\end{equation}
where $v_d$ is the velocity dispersion taken to be $220$~km/s, and $k_\textrm{iso}$ is a numerically determined normalization constant found by integrating the velocity profile over $d^3 \mathbf{v}$, with $v < v_{\textrm{esc}}$.  The canonical value of the galactic escape velocity $v_{\textrm{esc}}$ is  taken to be $544$~km/s. We denote the SHM VDF with these parameters as \textit{iso}. The local DM density is conventionally chosen to be $\rho_{DM,\odot}=0.3\,\mathrm{GeV }{\rm cm}^{-3}$; however, this value is not self-consistent with the assumed VDF parameters.

\subsection{Analytic fits to the observational VDFs}
Finally, for the sake of ease-of-use of our VDFs estimated from MW observations, we fit the VDFs  to the following functional form
\begin{equation}\label{Fit2}
f_\textrm{obs}(\mathbf{v}) \approx B( \zeta(\beta)- \zeta(\beta_{\textrm{max}})).
\end{equation}
where $\zeta(x) = (1+x)^k \exp[{-x^{(1-p)}}]$, $\beta = v^2/v_\star^2$ and $\beta_{\textrm{max}} = v_{\textrm{esc}}^2/v_\star^2$. Here $k\,,p$, $v_\star$  and $v_{\textrm{esc}}$  are fit parameters and $B$ is an appropriate normalization factor. The best fit parameters are given in Table \ref{TabFit} \, ; the fit parameters for the error envelopes are given in Appendix I. We show these fits for the \textit{obs} VDF  in Fig \ref{fig:Fit85}.  The analytic expression in eq.~(\ref{Fit2}) fits the \textit{obs} VDF to roughly 1\% accuracy over most of the velocity range, and is a better fit to the DM VDFs over other fitting forms proposed in the literature by \citet{Bhattacharjee:2012xm} and \citet{Mao2013}. A table of the VDFs along with the upper and lower uncertainty envelopes can be found at \href{https://github.com/sayanmandalcmu/darkmatterVDFs}{https://github.com/sayanmandalcmu/darkmatterVDFs}.

One might also consider the possibility of fitting the {\it obs} VDF with a Maxwellian profile with an alternate choice of the dispersion velocity parameter $v_d$. We have shown this as the best-fit Maxwellian in Fig~\ref{fig:Fit85} where $v_d = 217.3$~km~s$^{-1}$. From the figure we can clearly see that the fit is poor, as the Maxwellian VDF does not describe the MCMC determined VDF well, especially so in the high-velocity tail region.

\begin{table}[h]
\begin{tabular}{|l|c|c|c|c|}
\hline
VDF & $k$ & $p$ & $v_\star$ ($\mathrm{km\,s^{-1}}$) &   $v_{\textrm{esc}}$ ($\mathrm{km\,s^{-1}}$)  \\ \hline
B200-8.0-67 & $0.44$ & $-0.45$ & $262.99$ & $536.83$ \\ \hline
B220-8.5-67 (\textit{obs}) & $-2.48$ & $-1.69$ & $372.25$ & $475.00$ \\ \hline
\end{tabular}
\caption{Best fit parameters for eq. (\ref{Fit2}) for the empirically obtained VDFs. We recommend use of the {\it obs} VDF which is based on the IAU preferred LSR.}
\label{TabFit}
\end{table}

\section{Direct Detection Rate}

We will now examine the effect of the difference between the observationally determined (\textit{obs}) velocity profile and the canonical isothermal SHM profile that we have considered in the previous section on the dark matter direct detection rate. We will consider here only elastic scattering of DM with a target nucleus. Assuming isotropic scattering in the center-of-mass frame of the DM-nucleus system, the rate of direct detection signal events per unit recoil energy $(E_R)$, per unit detector mass, is given by \cite{Lewin:1996},
\begin{equation}
\label{eq:ddrate}
\frac{dR}{dE_R} = \frac{R_0}{E_0 r} \mathcal{I}(E_R)  F^2(E_R) \epsilon(E_R)
\end{equation}
The nominal rate $R_0$ is given by,
\begin{equation}
R_0 = \frac{320}{m_D m_T} \left( \frac{\sigma_0}{1 \textrm{ pb}} \right) \left( \frac{\rho_{DM,\odot}}{0.3 \textrm{ GeV/c$^2$}} \right) \left( \frac{v_0}{220 \textrm{ km/s}} \right) \textrm{ tru},
\end{equation}
where 1 tru is 1 count/kg/day. Here, the dark matter mass $m_D$ and target nucleus mass $m_T$ are expressed in GeV/c$^2$, $\sigma_0$ is the DM--target-nucleus cross-section, $\rho_{DM,\odot}$ is the local dark matter density and $v_0$ is a ``typical relative velocity'' parameter which is representative of the dark matter velocity in the detector rest frame. The factor of $v_0$ is introduced only for dimensional convenience and cancels out in the full expression for the rate.

An explanation of the other factors in  eq.~(\ref{eq:ddrate}) is in order here. In accordance with \cite{Lewin:1996}, $E_0= \frac{1}{2} m_D v_0^2$ is the characteristic recoil energy of the nucleus (typically of the order of a few keV for a 100 GeV dark matter particle) and $r = 4 m_D m_T/(m_D+ m_T)^2$. $F^2(E_R)$ is the nuclear form factor which is target dependent, here we use the Helm form-factor from \cite{Lewin:1996}. $\epsilon(E_R)$ is the detector efficiency as a function of recoil energy.

For spin-independent (and isospin independent) scattering the WIMP-nucleus cross-section $\sigma_0$ can be expressed in terms of the DM-\textit{nucleon} cross-section $\sigma_n$ by, $\sigma_0 = \left( \frac{\mu_{D,N}}{\mu_{D,n}} \right)^2  A^2 \sigma_n$, where $A$ is the nucleon number of the target, $\mu_{D,N}$ ($\mu_{D,n}$) is the reduced mass of the DM particle and the nucleus (nucleon). We will only consider spin-independent interactions in this work. The case of spin-dependent scattering is a trivial extension.

In eq.~(\ref{eq:ddrate}), the factor $\mathcal{I}(E_R)$ is a dimensionless velocity averaged integral given by,
\begin{equation}
\label{eq:Ifactor}
\mathcal{I}(E_R) =  \int_{v_r>v_{\textrm{min}}} \frac{v_0}{v_r} f(\mathbf{v_r} + \mathbf{v_e}) d^3 \mathbf{v_r}.
\end{equation}
Here, the integrand is evaluated over the relative velocity of a DM particle and the detector $\mathbf{v_r} = \mathbf{v_{\textrm{gal}}} + \mathbf{v_e}$, where $\mathbf{v_{\textrm{gal}}}$ is the DM velocity in the galactic rest frame and $\mathbf{v_e}$ is the earth's velocity. Since we will only focus on the time integrated recoil signal in this work~\footnote{Direct detection experiments can also be sensitive to diurnal/annual modulation effects in the recoil signal due to the Earth's rotation and revolution about the sun. The \textit{obs} VDF will also have an effect on the expected modulation signal that is different from that obtained by assuming the \textit{iso} VDF. We leave a study of this effect to future work.}, we take the earth's velocity to be a constant with magnitude $v_e = 240$~km/s. Demanding that the dark matter particle should have a large enough relative velocity to be able to cause a recoil energy $E_R$ in the detector gives us the lower bound on the relative velocity, $v_{\textrm{min}} = (2 E_R/(r m_D))^{1/2}$ . This introduces an explicit dependence of $\mathcal{I}$ on the recoil energy as well as the DM mass.

The velocity integral fully captures the way that the VDF of the dark matter affects the detection rate. Our VDFs by definition are cut-off above the galactic escape velocity. Thus, when $v_{\textrm{min}}$ exceeds $v_{\textrm{esc}} + v_e$, no recoils should occur. For a fixed DM mass, this implies that there is a maximum nuclear recoil energy given by $E_R^{\textrm{max}} =  \frac{1}{2} r m_D (v_{\textrm{esc}} + v_e )^2$ above which no signal is expected in the detector. We note that since the \textit{obs} VDF has a lower escape velocity compared to the isothermal VDF, we expect lower recoil energy cut-offs for a fixed DM mass when using the \textit{obs} velocity profile.

Conversely, for a fixed recoil energy, if $m_D$ is sufficiently large, then the velocity integral receives contributions from all possible relative velocities. In the absence of the factor $v_0/v_r$ in the integrand, the integral would evaluate to unity, independent of the velocity profile. Thus, for large DM masses, the distinction between the various VDFs arises mainly from the spread of DM relative velocities. However, for low DM masses, the support of the velocity integral shrinks and the rate becomes highly sensitive to the differences between the high velocity tails of the VDFs.

We can see from Fig.~\ref{fig:profile} that the \textit{obs} and isothermal VDFs have similar widths but the \textit{obs} velocity profile has a more suppressed high velocity tail. Thus, for a fixed recoil energy, we expect that the difference between the direct detection rates for the \textit{obs} and isothermal VDFs will be most dramatic for low dark matter masses.

\begin{figure}
\hspace*{-0.5cm}
\includegraphics[width=8.5cm]{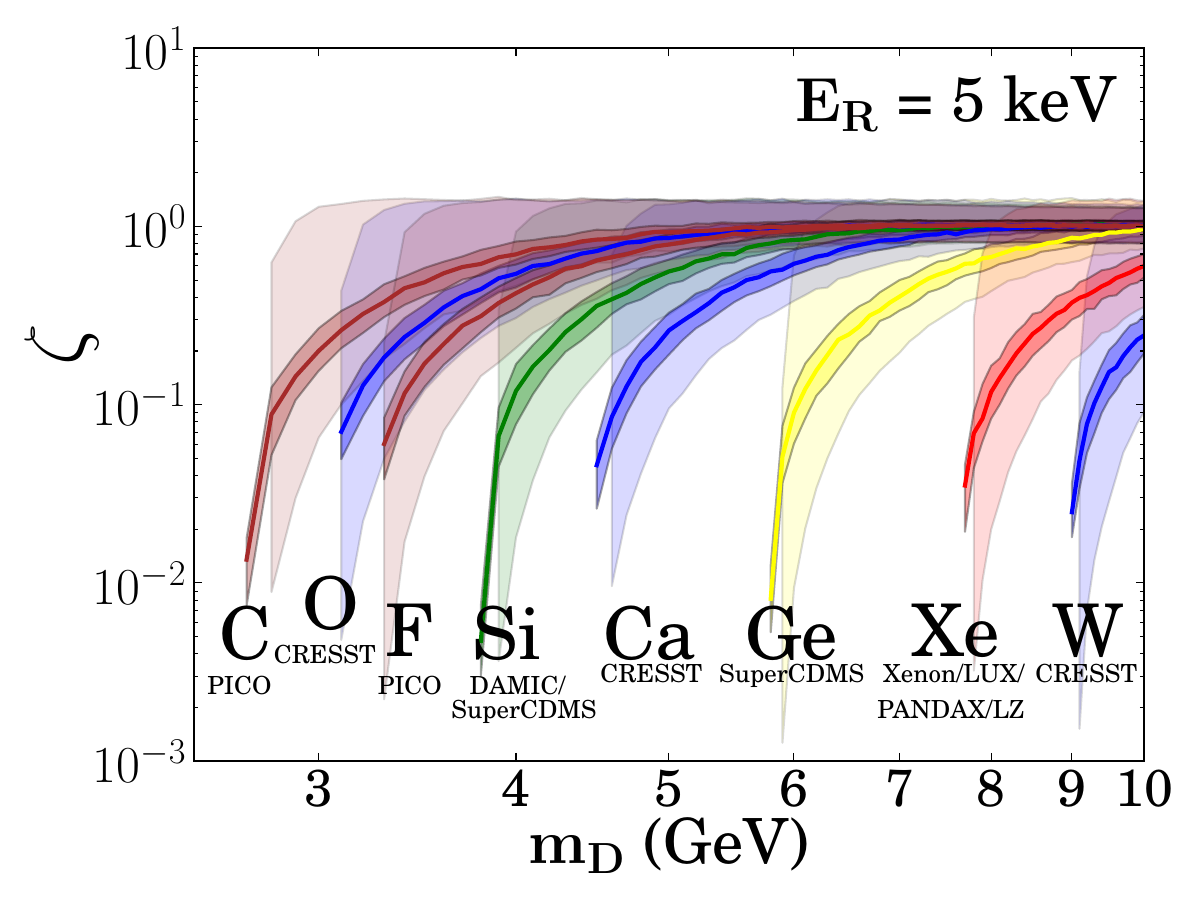}
\caption{Plot of $\zeta\equiv \mathcal{I}_{\textrm{obs}}/\mathcal{I}_{\textrm{iso}}$ the velocity integral ratio as a function of dark matter mass, for various target elements (solid lines) at a recoil energy of $E_R = 5$~keV.  The current and future experiments where these targets are in use are also given in the annotations. Significant deviations from the isothermal velocity profile are observed for low dark matter masses. The thick and thin envelopes indicate the uncertainty on $\zeta$ estimated by a propagation of the \textit{current} and \textit{third} velocity envelopes respectively, of the \textit{obs} VDF (see Fig.~\ref{fig:profile}).}
\label{fig:Ifactor_ratio_alldets_5kev}
\end{figure}
\begin{figure}
\hspace*{-0.5cm}
\includegraphics[width=8.5cm]{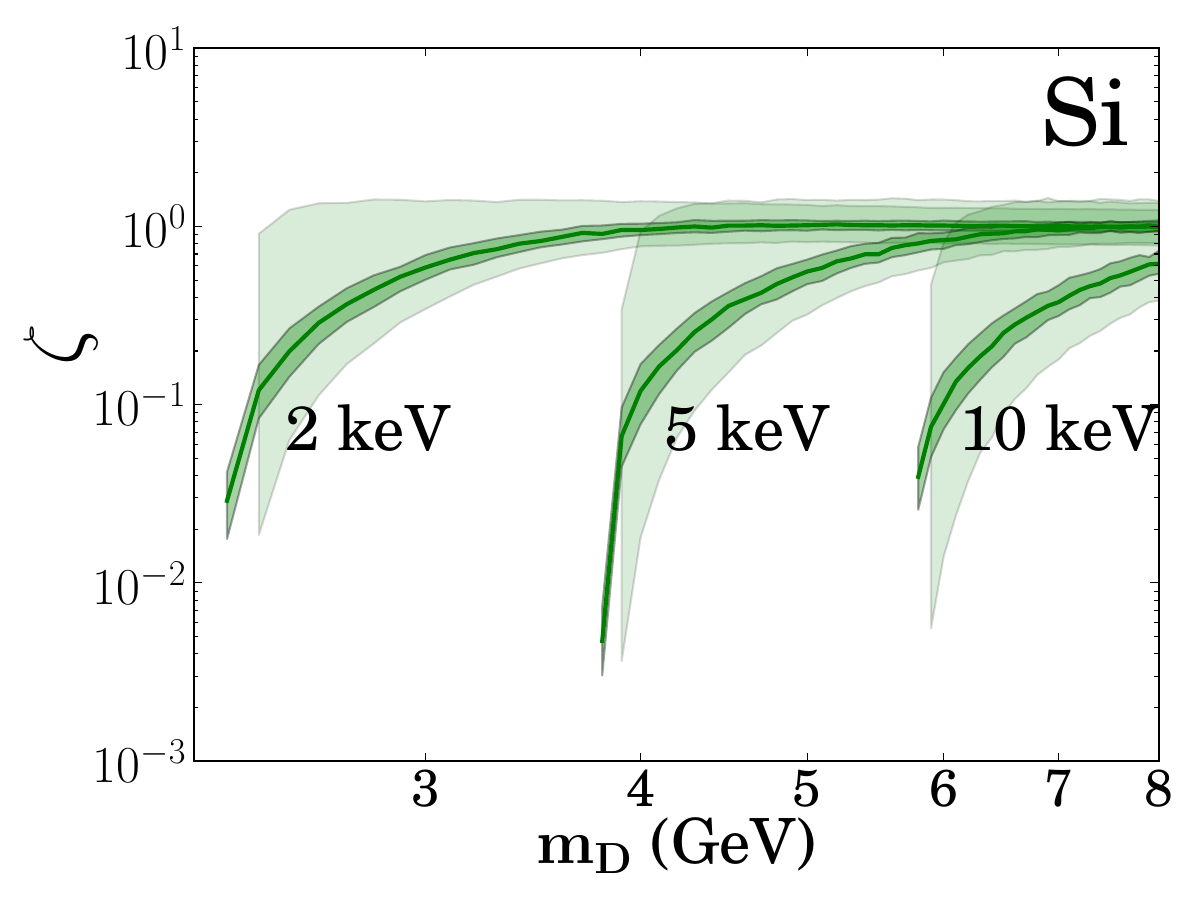}
\caption{Plot of $\zeta\equiv \mathcal{I}_{\textrm{obs}}/\mathcal{I}_{\textrm{iso}}$  the velocity integral ratio as a function of dark matter mass, for a Silicon target at various recoil energies (solid lines). The thick and thin envelopes indicate the uncertainty on $\zeta$ estimated by a propagation of the \textit{current} and \textit{third} velocity envelopes respectively, of the \textit{obs} VDF.}
\label{fig:Ifactor_ratio_Si_allrecoils}
\end{figure}

We define $\zeta \equiv \mathcal{I}_{\textrm{obs}}/\mathcal{I}_{\textrm{iso}}$ as the ratio of $\mathcal{I}$ factors for the fiducial \textit{obs} velocity profile and the isothermal profile. In Fig.~\ref{fig:Ifactor_ratio_alldets_5kev}, we plot $\zeta$  as a function of the dark matter mass,  for several different target nuclei, for a fixed  recoil energy $E_R = 5$~keV. For low DM masses compared to the target mass, $\left ( v_{\textrm{min}} \rightarrow  ( E_R m_T/2 (m_D)^2)^{1/2}  \right)$. Thus, for a given target and a fixed recoil energy, the $\mathcal{I}$ factor is sensitive to the tail of the VDF. The tail of the \textit{obs} VDF falls below that of the isothermal profile at a galactic velocity $v_{\textrm{dev}} \simeq 330$~km~s$^{-1}$ (see Fig.~\ref{fig:profile}). Hence, the $\mathcal{I}$ ratio drops below unity for low DM masses,  $ m_D  \rightarrow  ( E_R m_T/2 (v_{\textrm{dev}} + v_e )^2)^{1/2} $.

In Fig.~\ref{fig:Ifactor_ratio_Si_allrecoils}, we show how $\zeta$ varies as a function of dark matter mass for a Si target at various recoil energies. We can see from the figure that at higher recoil energies the deviation of $\zeta$ from unity occurs for correspondingly higher values of the dark matter mass. The error bands in both Figs.~\ref{fig:Ifactor_ratio_alldets_5kev} and ~\ref{fig:Ifactor_ratio_Si_allrecoils} reflect the propagation of the \textit{current} and \textit{third} VDF uncertainty envelopes of Fig.~\ref{fig:profile} into the corresponding velocity integral, $\mathcal{I}_{\textrm{obs}}$.

For a given recoil energy it thus seems that the $\mathcal{I}$-factors can vary by several orders of magnitude for low dark matter masses. We would thus expect a large change in the total direct detection signal rates and consequently the exclusion bounds for low dark matter masses if we used the \textit{obs} VDF rather than the canonical isothermal profile. However, as we shall see next this expectation is tempered by the fact that at low recoil energies we actually get a small contribution to the overall rate due to the low detector efficiency at these energies.

\section{Results}
\label{sec:results}
\begin{figure}[ht]
\includegraphics[width=8cm]{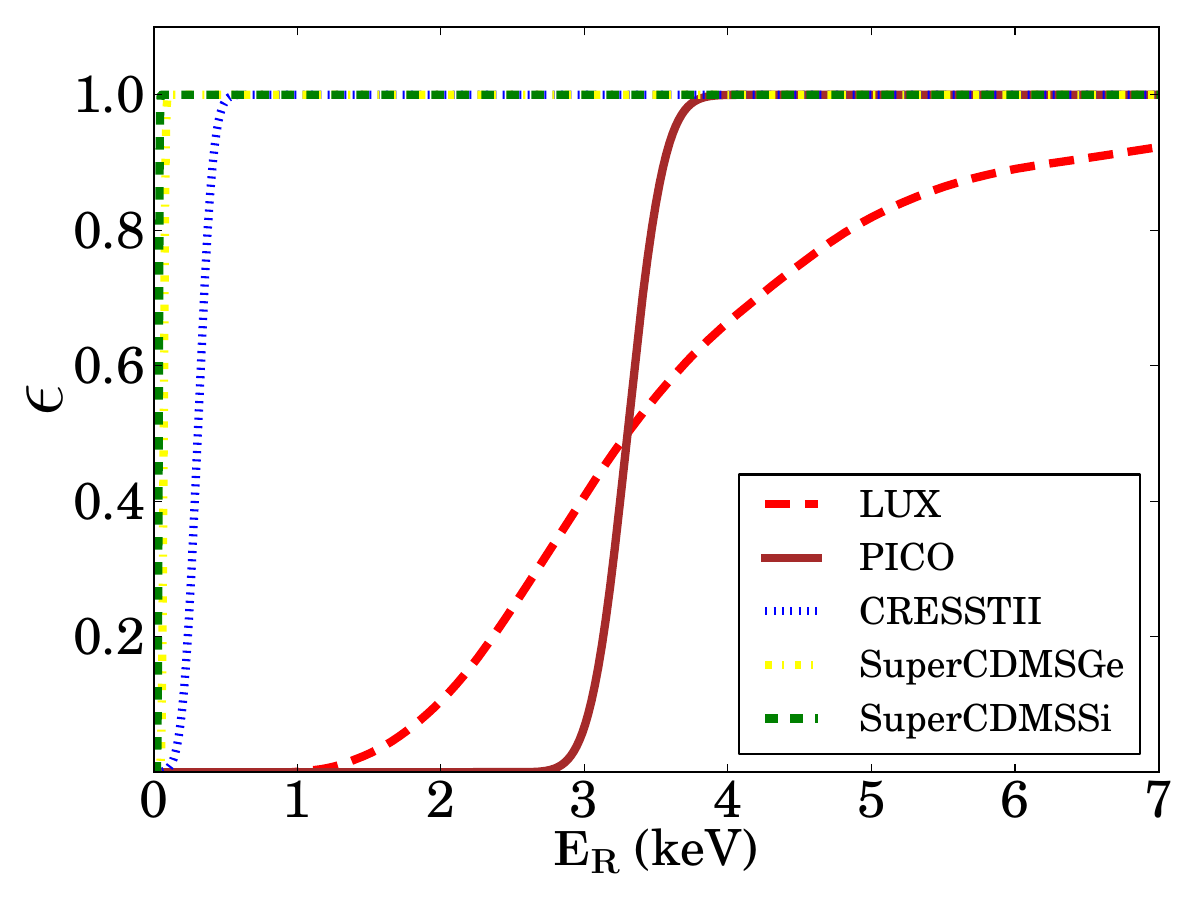}
\caption{Plot of detector efficiencies as a function of recoil energy for LUX~\cite{Akerib:2015rjg}, PICO~\cite{Amole:2017dex}, CRESSTII~\cite{Angloher:2015ewa} and the proposed SuperCDMS Si and Ge High Voltage detectors~\cite{Agnese:2016cpb,BasuThakur:2014oma, Pepin:2016ynh}.}
\label{fig:det_effs}
\end{figure}

In this section we would like to see whether our expectation of the strong sensitivity of the recoil rate to the VDF continues to hold upon including detector efficiency effects. In Fig.~\ref{fig:det_effs} we plot the detector efficiency for several existing experiments CRESSTII~\cite{Angloher:2015ewa}, LUX~\cite{Akerib:2015rjg}, and PICO~\cite{Amole:2017dex}. We also show the efficiencies for the proposed SuperCDMS experiment using published projections for the Silicon and Germanium High Voltage (HV) detectors~\cite{Agnese:2016cpb,BasuThakur:2014oma, Pepin:2016ynh}. Each efficiency function can be characterized by a threshold energy, $(E_T)$ which can be defined as the recoil energy for which the efficiency drops to $50\%$ of maximum efficiency, and the width of the detector response near the threshold (conventionally defined by parameterizing the response as an error-function, see discussion on detector specifications in Appendix II).

\begin{figure}[ht]
\includegraphics[width=8cm]{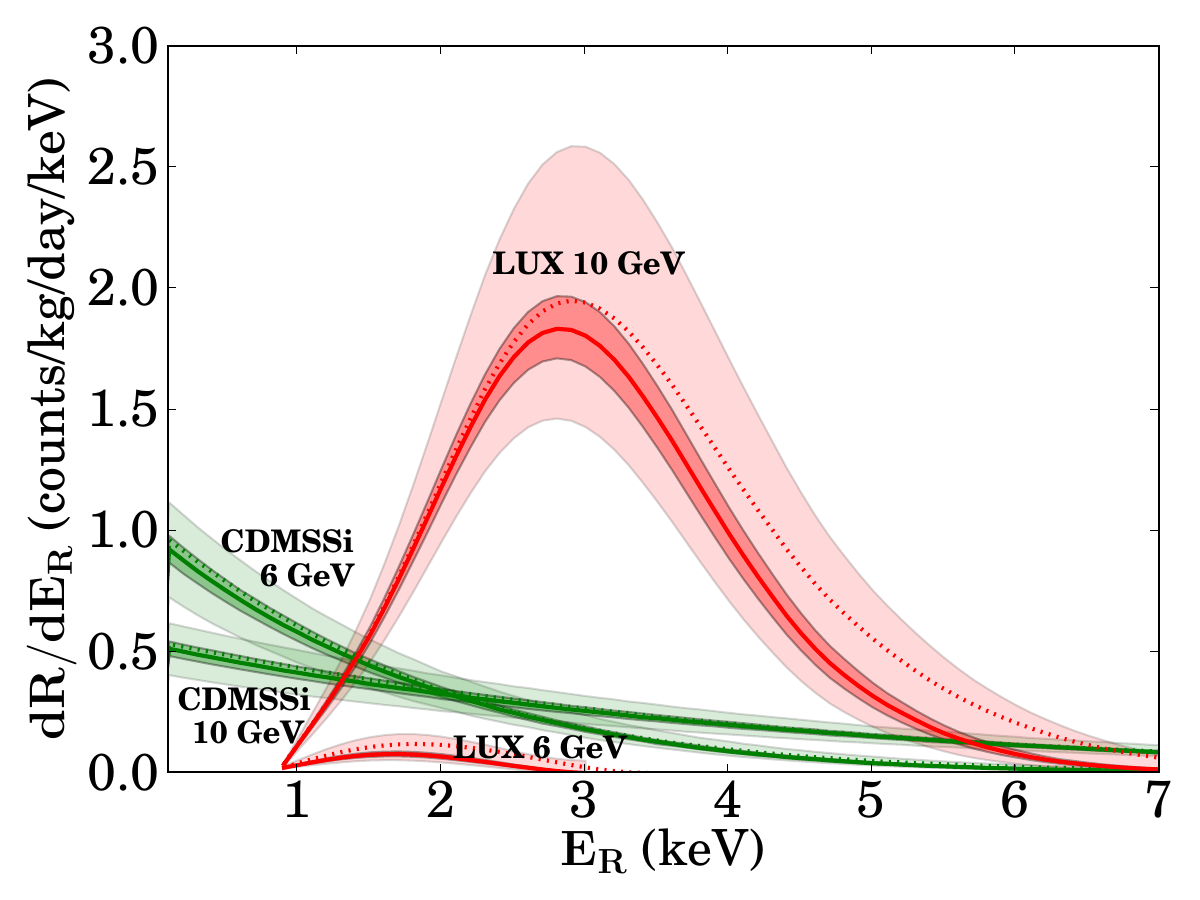}
\caption{Plot of the recoil energy spectrum (solid lines) for dark matter particles with masses 6 GeV and 10 GeV scattering off of a Si target (SuperCDMS) and Xe target (LUX) for the \textit{obs} velocity profile. The corresponding spectra with the isothermal (SHM) VDF are shown with dotted lines. Here, we have assumed a WIMP-nucleon scattering cross-section $\sigma_n = 10^{-40}$~cm$^2$. The thick and thin envelopes indicate the uncertainty on the recoil spectra  estimated by a propagation of the \textit{current} and \textit{third} velocity envelopes respectively, of the \textit{obs} VDF.}
\label{fig:recoil_spectrum}
\end{figure}

Using these efficiencies, we can now compute the recoil energy spectra for different detectors for the \textit{obs} and SHM VDFs using eq.~(\ref{eq:ddrate}). Note that we need to choose the appropriate value of the local DM density corresponding to the VDF that we are using, which changes the overall normalization of the rate for different VDFs. It is the strength of our present approach, as detailed in previous sections, that we have self-consistent pairs of local DM density and VDFs.
In Fig.~\ref{fig:recoil_spectrum} we plot the recoil spectra due to DM-nucleus interactions for the SuperCDMS Silicon detector and the LUX detector for DM masses of 6 GeV and 10 GeV, assuming a WIMP-nucleon cross-section $\sigma_n = 10^{-40}$~cm$^2$.

We note a few interesting features of the recoil spectra. The high energy tail of the recoil spectra is sensitive to the tail of the VDFs. However, at low recoil energies the shape of the recoil spectrum is determined by the threshold energy and efficiency of the detector near the threshold. We note that each detector has a minimum dark matter mass below which no events are seen in the detector. This minimum mass is given by
$\left ( m^{\text{min}}_D  \simeq  ( E_R^\textrm{min} m_T /2 (v_{\textrm{esc}} + v_e )^2)^{1/2}  \right)$, where $v_{\textrm{esc}}$ is the local escape velocity and $E_R^\textrm{min}$ is the minimum deposited energy that can be detected. Since CRESST and SuperCDMS are low threshold experiments, they are sensitive to much lower recoil energies and hence to much lower dark matter masses.

\begin{figure}[ht]
 \includegraphics[width=8cm]{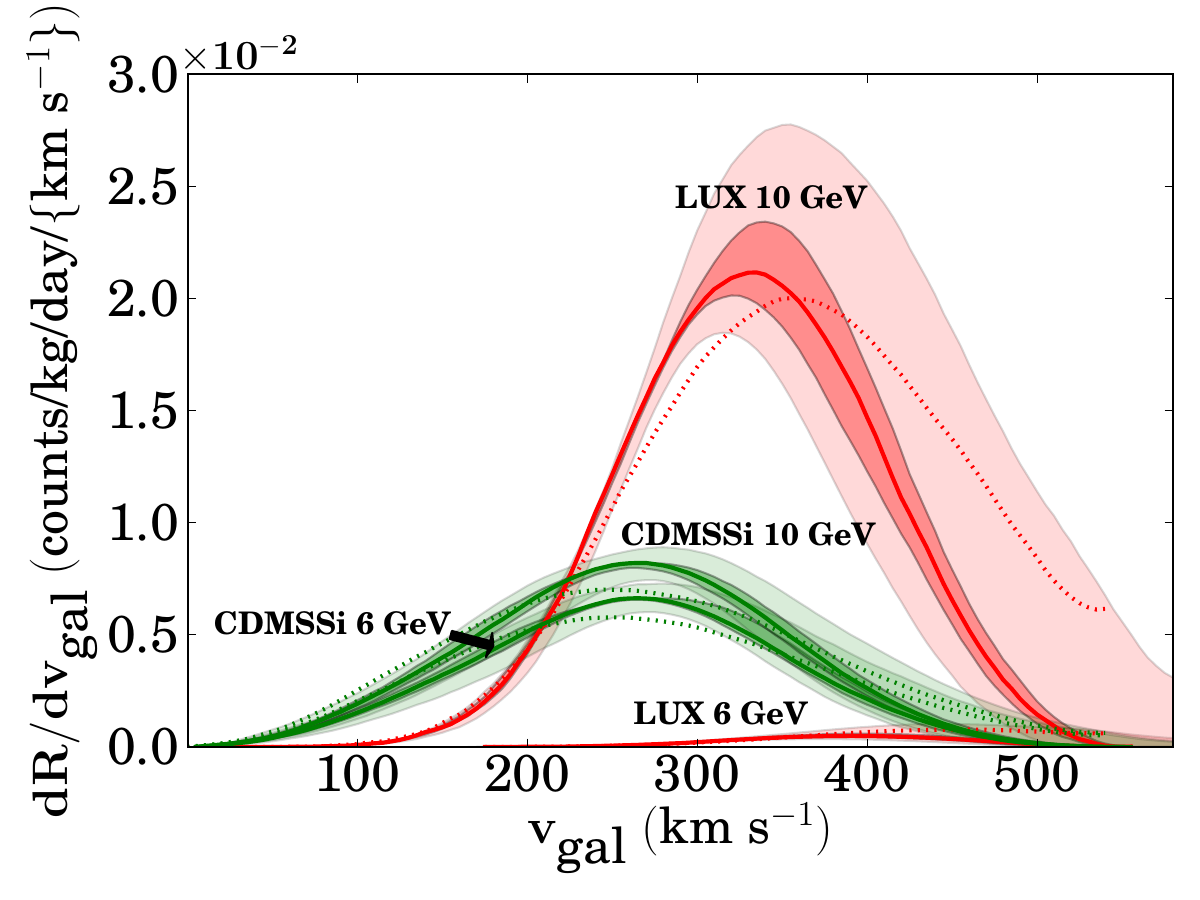}
\caption{Plot of the contribution to total direct detection rate (solid lines) for a 6 GeV and 10 GeV dark matter particle scattering in LUX and SuperCDMS (Si) as a function of $\textrm{v}_\textrm{gal}$ (the speed of DM in the galactic rest frame). We have assumed a WIMP-nucleon scattering cross-section $\sigma_n = 10^{-40}$~cm$^2$.  The thick and thin envelopes indicate the uncertainty on the recoil rate estimated by a propagation of the \textit{current} and \textit{third} velocity envelopes respectively, of the \textit{obs} VDF. The dashed lines show the corresponding rate assuming the \textit{iso} VDF.}
\label{fig:rate_vgal}
\end{figure}

In Fig.~\ref{fig:rate_vgal} we show the differential recoil rate $dR/dv_\textrm{gal}$ as a function of the magnitude of the dark matter velocity in the galactic rest frame $(\mathbf{v}_\textrm{gal} = \mathbf{v_r} + \mathbf{v_e})$ for DM particles scattering in LUX and SuperCDMS with the same benchmark cross-section. Comparing this plot with the VDFs in Fig.~\ref{fig:profile} allows us to visualize the contribution to the total recoil signal for different ranges of DM galactic velocities. From the figure, we can clearly see the recoils of low mass DM probe the high velocity tail of the VDFs.

\begin{figure}[ht]
 \includegraphics[width=8cm]{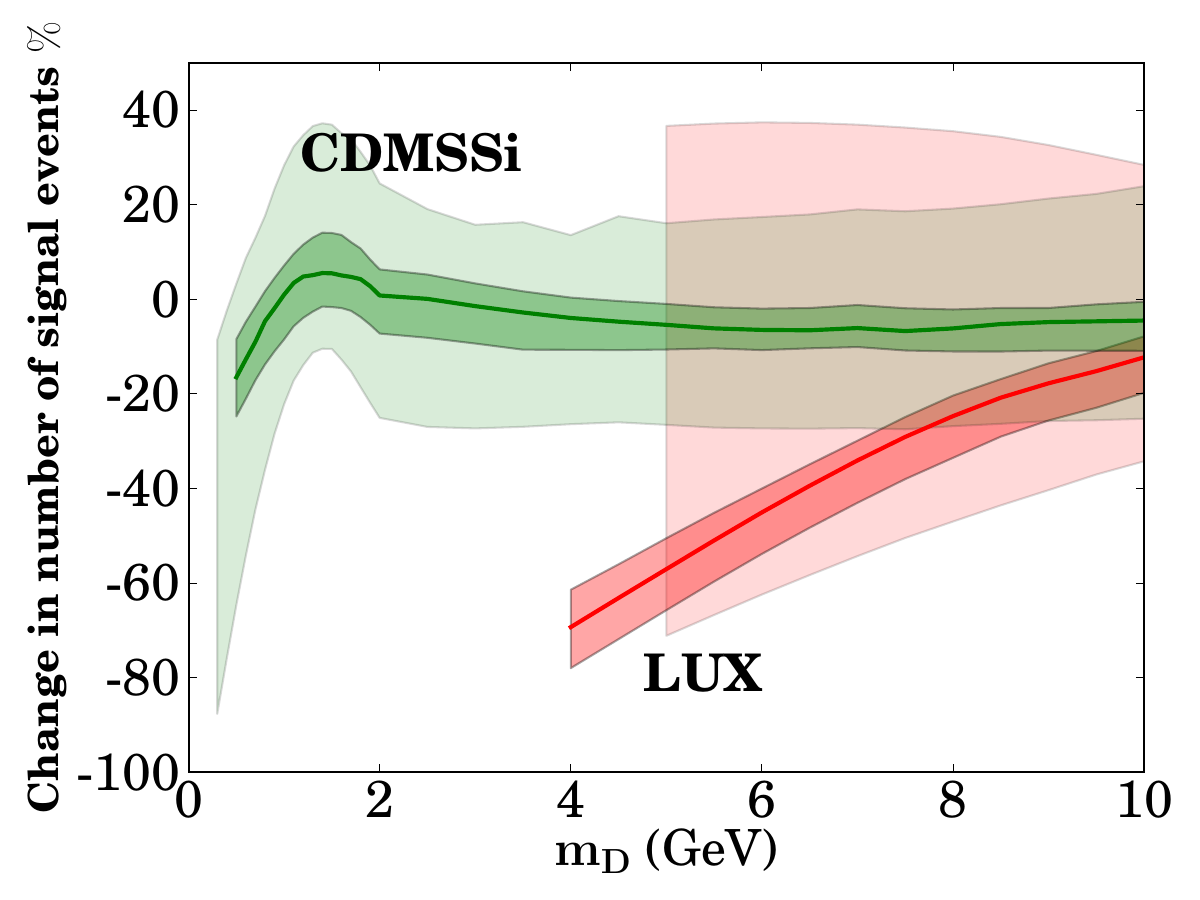}
\caption{Plot of the change (in percent) of the expected number of signal events assuming the \textit{obs} VDF as opposed to the canonical SHM VDF, as a function of dark matter mass for the SuperCDMS (Si) and LUX detectors (solid lines). The thick and thin envelopes indicate the uncertainty on this difference estimated by a propagation of the \textit{current} and \textit{third} velocity envelopes respectively, of the \textit{obs} VDF.}
\label{fig:sigratio}
\end{figure}

Finally, we can examine the difference between the fully integrated signal rates for the \textit{obs} and isothermal VDFs. In Fig.~\ref{fig:sigratio}, we plot as a function of dark matter mass the relative difference (in percent) of the expected number of signal events using the observationally determined VDFs and the expected number obtained by using the SHM.

To derive a realistic upper bound on the DM-nucleon cross-section ($\sigma_n$) from experiment requires an understanding of the detector background recoil spectra shape and corresponding systematic uncertainties, understanding of the detector efficiency and threshold as well as knowledge of the detector exposure. A detailed study of these effects is typically possible only within each experimental collaboration. Here, we have estimated a simple bound on $\sigma_n$ for the LUX, PICO and CRESSTII~\cite{Angloher:2015ewa} experiments. We have also computed an expected exclusion limit for the SuperCDMS Silicon and Germanium high-voltage detectors.

Our procedure was as follows: We first assumed detector exposures and efficiencies for several detectors based on their published results where applicable. These exposures and efficiencies are summarized in the appendix, along with the references from which they were obtained. For all experiments we assumed a background rate of 1~dru~$\equiv$~1~count/keV/kg/day which is constant over recoil energies from 0-200 keV. We then calculated a simple median $90\%$~CL estimated bound on the dark matter nucleon cross-section $\sigma_n$ by doing a simple counting pseudo-experiment over the entire recoil energy range from 0-200~keV, assuming that only background is observed.

The upper end of the recoil energy range of 200 keV is an arbitrary choice. This value is much higher than the maximum recoil energy expected for low DM masses. We are unable to perform a profile likelihood analysis \cite{Cowan2010} which would take into account shape differences between the signal and background without a detailed understanding of the detector backgrounds, but a detailed experimental analysis would optimize the choice of this upper cut-off for every candidate DM mass. We assumed a simple flat background spectrum, and therefore our choice of the upper recoil energy cut-off of 200 keV added a fixed amount of background that could realistically be reduced.

\begin{figure}[ht]
\hspace*{-0.5cm}
 \includegraphics[width=9cm, height= 10cm]{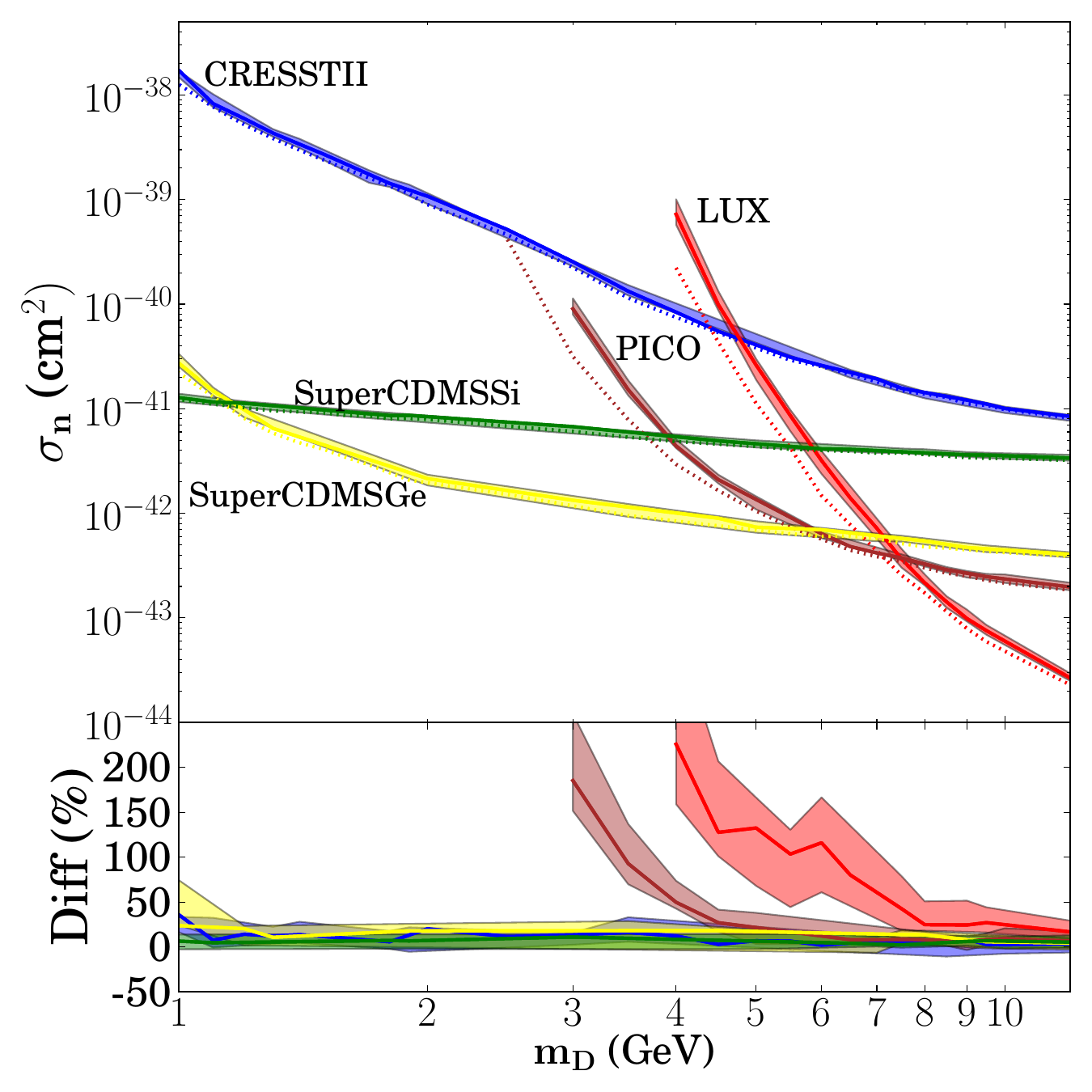}
\caption{Estimated median $90\%$~CL upper-bound on spin-independent DM nucleon scattering cross-section $\sigma_n$ for several experiments. The top panel shows the estimated bound using the \textit{obs} VDF (solid lines) along with the \textit{third} uncertainty envelope. The thicker \textit{current} uncertainty envelope is omitted for clarity. The estimated bounds for the isothermal VDF are shown with dashed lines. The differences between the \textit{obs} and isothermal bounds are shown in the lower panel. We can see that the difference is most significant for masses close to the detector threshold mass $m^{\text{min}}_D$. This difference is significant when VDF errors improve to \textit{third} errors (shown); with \textit{current} errors  (not shown) the difference cannot be resolved.}
\label{fig:estimated_bound}
\end{figure}

We plot our estimated upper-bound on $\sigma_n$ for the \textit{obs} and SHM velocity profiles for the selected experiments in the upper panel of Fig.~\ref{fig:estimated_bound}. Note that it is important to use appropriate choice of local DM density corresponding to the selected VDF when computing the exclusion curves. We also show the uncertainty bands around the exclusion curves for the {\it obs} VDF corresponding to a straight forward propagation of the {\it third} VDF uncertainty envelope from Fig.~\ref{fig:profile}\,.
The lower panel of Fig.~\ref{fig:estimated_bound} shows the percentage difference between the exclusion bounds set by assuming the {\it obs} and SHM local DM phase-space.

There are several interesting features in Fig.~\ref{fig:estimated_bound}\,.
We note that there is a few percent normalization difference in the bounds for the different velocity profiles,  which is apparent at high DM masses, due  to the difference in the  local DM density for each profile choice. In addition to the density effect, the estimated bounds on $\sigma_n$ differ in shape, at low dark matter masses, due to impact of the different velocity profiles .
This difference is most stark at the threshold mass  $m^{\text{min}}_D$ of each detector where  $m^{\text{min}}_D \simeq (E_T m_T/2)^{1/2} / (v_{\rm esc}+v_e)$ and $E_T$ is the detector threshold energy. The exact value of $m^{\text{min}}_D$, as well as the shape of the exclusion curve near $m^{\text{min}}_D$, depend on the tails of the VDFs.
Additionally, they also depend on  the detector sensitivity near the detector threshold, which is different for each detector.

For example, taking the case for LUX, which has $m^{\text{min}}_D \simeq 4$ GeV, the  exclusion limit obtained by assuming the {\it obs} VDF profile as opposed to the SHM VDF yields  $> 200$\% difference near threshold, i.e.  the constraints are weakened by a factor of three. Similarly, for PICO, $m^{\text{min}}_D \simeq 3$ GeV, and the exclusion limit differs by $> 150$\% near threshold. For the lower threshold experiments, such as CRESSTII and the near future SuperCDMS experiments,  $m^{\text{min}}_D < 1$ GeV, and the difference in the exclusions are $\lesssim 50$\%.

In order to judge whether these differences in the exclusion curves when using the \textit{obs} versus \textit{iso} VDFs are significant, we need to consider the uncertainty on the \textit{obs} VDF profile. Given the \textit{current} errors, the difference between the central values of the exclusion curves is well within the VDF uncertainty. However, assuming the same central values of the VDFs but with errors reduced (by future astrophysical measurements) to \textit{third} errors, we find that for LUX and PICO, this difference could be $\sim 5\sigma$ significant.

In a direct detection experiment, there are four main sources of systematic errors : (i) astrophysical uncertainties on the local DM density and the VDF (ii) detector response uncertainty, (iii) uncertainty of the nuclear form factors and (iv)  uncertainty on the detector background. In this work, we have discussed the uncertainty on the local DM density and VDF and argued that the VDF uncertainty is the dominant astrophysical unknown. To assess the impact of other uncertainties on the DM exclusion bound requires a careful understanding of the detector and is beyond the scope of this work.

To get an idea of the relative importance of the astrophysical vs detector uncertainties, we use the published uncertainties on the LUX exclusion \cite{Akerib:2015rjg} as a benchmark. An examination of their bounds indicate $\sim 50\%$ detector related uncertainties, at all candidate DM masses. Note, that they do not include astrophysical uncertainties in their exclusion limits. In contrast, expected uncertainties in the mean DM exclusion curve due to {\it third} errors on the {\it obs} VDFs are $\sim 30\%$ and, hence, it is expected to be subdominant to detector systematics once precision astrophysics results are used to determine the DM VDF.

Given that the application of the central {\it obs} VDF results in a systematic deviation of up to $200\%$ in the mean DM exclusion curve, along with an estimate of the combined expected uncertainty from astrophysics as well the currently known experimental systematics of $\sim 60$\%, it is clearly important  to use the best available observationally determined VDF when presenting the results of DM direct detection experiments.

\section{Discussion and Conclusions}
We have constructed an observationally driven determination of the local DM density and velocity distribution and used this to interpret the null results of DM direct detection experiments. Milky Way astrophysical data is poised for unprecedented precision measurements with the release of GAIA data and can be used to get precise estimates of the MW (and in particular the local) DM phase-space distribution. This approach goes beyond the simplistic, and incorrect, isothermal VDF in the so-called SHM of the MW. Previous attempts to go beyond the SHM have mainly relied on simulations of a MW-{\it like} DM halo. These simulations, although state-of-the-art, are not guaranteed to describe exactly our MW. Despite the promise shown by these simulation results,  it is prudent to use observationally inferred DM phase-space distributions. Our work is unique among other attempts in modelling the local DM phase-space in that it goes beyond just local kinematic data (prone to large DM-VM degeneracies) and uses rotation curve data up to $\sim 200$ kpc for a full Bayesian reconstruction of the MW mass model and its corresponding self-consistent local DM density and VDF.

The mean observational VDF that we have determined differs from the SHM isothermal VDF in that it has a lower escape velocity and is significantly non-Maxwellian especially at the tails. Current RC data has large error bars for the majority of radial bins and results in a large {\it current} error band around the mean VDF. However, even with this large error bar, the {\it obs} VDF clearly differs from the isothermal VDF, in particular, at the high velocity tail. A similar determination of the DM VDF from the GAIA astrometric observations \footnote{https://www.cosmos.esa.int/web/gaia/dr2} are expected to reduce the error on VDF significantly, thus potentially differentiating the {\it obs} VDF from the isothermal VDF over most of the DM velocity range.

Low mass ($\sim$ few GeV) DM has received considerable interest due to claims of a long-standing detection of an annual modulation signal by DAMA \cite{Bernabei:2018yyw}, as well as claims of excess events seen in other direct detection experiments such as CDMS-II~\cite{Ahmed:2009zw}, COGENT~\cite{Aalseth:2010vx} and CRESST-II~\cite{Angloher:2011uu}. In addition, the observed excess of gamma rays from the galactic center could also be explained by the annihilation of a low mass DM species~\cite{Daylan:2014rsa, TheFermi-LAT:2017vmf}. Although the DM origin of these anomalies is far from certain, these results are indicative of the need to precisely interpret the results of direct detection experiments for low DM masses.

In this work, we have shown that the difference between the observationally determined VDFs and the conventionally used isothermal VDF can yield very different interpretations of direct detection experiments for low candidate DM masses. Using the right DM VDF becomes especially pertinent when using VDFs inferred from future measurements from the GAIA telescope (where for simplicity we assume in this work that the mean {\it obs} VDF remains the same whereas the uncertainty around the mean reduces). For example, in such a scenario, for DM experiments like LUX or PICO, the DM exclusion limit using the {\it obs} DM VDF is expected to deviate by up to $5\sigma$ from the limit inferred from the SHM VDF at the detector threshold DM mass sensitivity. For future low threshold experiments, like SuperCDMS, accurate knowledge of the shape of the detector response is crucial to compute the impact of using the observationally determined DM VDF.

We emphasize that it is imperative that DM experiments use the best observationally estimated DM VDF when setting exclusion limits. For this purpose we have provided an accurate analytic fit to the {\it obs} VDF given in eq.~\eqref{Fit2} and a github link to the tables of the actual VDFs.

\section{Acknowledgments}

We would like to thank R Catena, Hamish Silverwood and Pierro Ullio for useful discussions. SM~(TIFR) also acknowledges useful conversations with participants at the MIAPP workshop on DM physics. SM~(TIFR) and VR would like to thank the Abdus Salam International Center for Theoretical Physics where a part of this work was completed. VR is supported by a DST-SERB Early Career Research Award (ECR/2017/000040) and an IITB-IRCC seed grant.

\appendix
\section{Appendix I: Fit to VDF envelopes}
\label{sec:appendix1}
We provide the best fit parameters for the upper and lower edges of the error envelope for the VDFs shown in Fig.~\ref{fig:profile} in Tables \ref{TabFit1} \& \ref{TabFit2}. Note that we have used the same parameterizations as eq.~\ref{Fit2} although the fitting function is ideal for the mean VDF.

\begin{table}[h]
\begin{tabular}{|l|c|c|c|c|}
\hline
VDF & $k$ & $p$ & $v_\star$ ($\mathrm{km\,s^{-1}}$) & $v_{\textrm{esc}}$ ($\mathrm{km\,s^{-1}}$) \\ \hline
B200-8.0-67 & $-0.22$ & $-0.26$ & $275.86$ & $500.32$ \\ \hline
B220-8.5-67 (\textit{obs}) & $-4.89$ & $-4.25$ & $455.03$  & $444.41$ \\ \hline
\end{tabular}
\caption{Best fit parameters for eq. (\ref{Fit2}) for the upper envelope of the empirically obtained VDFs.}
\label{TabFit1}
\end{table}

\begin{table}[h]
\begin{tabular}{|l|c|c|c|c|}
\hline
VDF & $k$ & $p$ & $v_\star$ ($\mathrm{km\,s^{-1}}$) & $v_{\textrm{esc}}$ ($\mathrm{km\,s^{-1}}$)\\ \hline
B200-8.0-67 & $0.52$ & $-0.99$ & $288.29$  &  $601.05$ \\ \hline
B220-8.5-67 (\textit{obs}) & $-0.99$ & $-1.84$ & $332.32$ & $518.57$  \\ \hline
\end{tabular}
\caption{Best fit parameters for eq. (\ref{Fit2}) for the lower envelope of the empirically obtained VDFs. }
\label{TabFit2}
\end{table}

\section{Appendix II: Detector specifications}
\label{sec:appendix2}
For experiments other than LUX, the detector response function was taken to have the parametric form,
\begin{equation}
\epsilon(E_R) = \frac{1}{2} \left ( 1 + \textrm{Erf} \left( \frac{E_R - E_T}{\sqrt{2} \sigma } \right) \right )
\end{equation}
in terms of the error-function. Here $E_T$ is the detector threshold energy at which the efficiency drops to 50\% and $\sigma$ is the width of the efficiency near threshold. For the LUX experiment we used the efficiency curve given in Fig. 1 of ref.~\cite{Akerib:2015rjg} with a hard low energy cut-off of 1.1~keV. We present for completeness in Table~\ref{tab:deteffs} a listing of the detector properties for the experiments used in the results presented in the main text. We also list the references from which these specifications were extracted.

\begin{table*}[t]

\centering
\begin{center}
\begin{tabular}{|x{1cm}|x{2.5cm}|x{1.8cm}|x{1.8cm}|x{1.8cm}|x{1.5cm}|x{1.5cm}|x{1.5cm}|}
  \hline
  S. No. & Detector Name & Target material & $E_T$  & $\sigma$  &   Current Exposure (kg-days) & Future Exposure (kg-days)   & Ref \\ \hline \hline
  1.  & LUX &  Xe & (see text) & (see text) &   $1.4\times10^4$ & - &    \cite{Akerib:2015rjg} \\  \hline
  2.  & CRESST II & CaWO$_4$   &  0.31 keV & 0.08 keV &  52  & - &    \cite{Angloher:2015ewa} \\  \hline
  3.  & PICO & C$_3$F$_8$ & 3.3 keV & 0.2 keV &  1167  & - &   \cite{Amole:2017dex} \\  \hline \hline
  4.  & SuperCDMS HV Si & Si  & 35 eV & 5~eV & -  & 9.6 &    \cite{Agnese:2016cpb} \\  \hline
  5.  & SuperCDMS HV Ge & Ge & 70 eV & 10~eV & -  & 44 &   \cite{Agnese:2016cpb} \\  \hline
  \hline
\end{tabular}
\end{center}
\caption{Table of detector properties used for the results presented in the main text. The parameters $E_T$ and $\sigma$ are used to construct an analytic approximation to the detector efficiency curve. For the proposed SuperCDMS High Voltage experiments, we have used the estimated future exposures shown in the table, whereas we have used current exposures for other experiments.}
\label{tab:deteffs}
\end{table*}

\bibliography{direct_det}

\end{document}